\documentclass[final,5p,times,twocolumn]{elsarticle}

\usepackage{graphicx}

\usepackage{amssymb}
\usepackage{amsthm}
\usepackage{amsmath}

\journal{Physica B}

\begin{document}

\begin{frontmatter}



\title{Determination of screened Coulomb repulsion energies in organic molecular crystals: A real space approach}


\author[label1]{Laura Cano-Cort\'es}
\author[label2]{Andreas Dolfen}
\author[label1]{Jaime Merino}
\author[label2]{Erik Koch}

\address[label1]{Departamento de F\'isica Te\'orica de la Materia Condensada,
Universidad Aut\'onoma de Madrid, Madrid 28049, Spain}
\address[label2]{German Research School for Simulation Sciences, Forschungszentrum J\"ulich and RWTH Aachen
University, 52425 J\"ulich, Germany}

\begin{abstract}

We present a general method for determining screened Coulomb parameters in molecular assemblies, in particular organic molecular crystals. This allows us to calculate the interaction parameters used in a generalized Hubbard model description of correlated organic materials. In such a model only the electrons in levels close to the Fermi level are included explicitly, while the effect of all other electrons is included as a renormalization of the model parameters. For the Coulomb integrals this renormalization is mainly due to screening. For molecular materials we can split the screening into intra- and inter-molecular screening. Here we demonstrate how the inter-molecular screening can be calculated by modeling the molecules by distributed point-polarizabilities and solving the resulting self-consistent electrostatic screening problem in real space. For the example of the quasi one-dimensional molecular metal TTF-TCNQ we demonstrate that the method gives remarkably accurate results.
\end{abstract}


\end{frontmatter}



\section{Introduction} 
\label{sec:intro}

Low dimensional organic molecular compounds based on the BEDT-TTF, TTF and TCNQ molecules are narrow band systems,
where the Coulomb repulsion is large compared to the bandwidth. This leads to many-body effects which are manifest,
e.g., in the spin susceptibility \cite{Torrance} or angular-resolved photoemission experiments
\cite{Jerome,Ito05,Sing}. While theoretical works suggest the importance of Coulomb parameters in these materials
\cite{Torrance,Mazumdar}, only rough theoretical estimates of these parameters do exist \cite{Hubbard}. The use of
modern techniques, such as density-functional theory (DFT), gives accurate results of the parameters for individual
molecules \cite{TTF-TCNQ,Scriven}, improving previous quantum chemistry calculations \cite{Castet,Mori}. However, there is a lack of accurate theoretical estimates of the Coulomb repulsion screened inside the crystal. These are needed for a more realistic description of organic molecular compounds.

In the present paper we introduce a systematic method for calculating these Coulomb parameters in low-dimensional organic molecular crystals. In Sec.~\ref{sec:model} we construct the minimal model and give details of the general formalism. This is applied to TTF-TCNQ salts in Sec.~\ref{sec:submolecular}, where we discuss the distributed-dipole approach. In Sec.~\ref{sec:screening} we give our results for Coulomb parameters.

\section{Model and Formalism}
\label{sec:model}
The minimal model generally used to describe the low-energy electronic properties of low-dimensional organic systems
is the Hubbard model \cite{Anderson}, where transfer integrals between neighboring sites and on-site Coulomb
interactions are taken into account. Such a simple Hubbard model is, however, usually not sufficient to describe
strongly correlated organics, as long-range Coulomb repulsion energies can rarely be neglected \cite{Hubbard,Horsch}. In the generalized model
\begin{eqnarray}
 \label{eq:HubbardModel}
 H=-t\!\!\sum_{ \langle ij\rangle,\sigma}
 (c^{\dagger}_{i,\sigma} c_{j,\sigma} + h.c.)
 + U\!\sum_i n_{i\uparrow} n_{i\downarrow}
 + \sum_{i,j} V_{ij} n_i n_j,
\end{eqnarray}
we include hopping integrals between nearest-neighbors, $t$, on-site Coulomb energies, $U$, and longer-range Coulomb interactions, $V_{ij}$. While transfer integrals can be obtained from {\em ab initio} DFT calculations \cite{TTF-TCNQ,Valenti,Nakamura}, Coulomb parameters are more difficult to obtain, as they include the screening correction due to different high-energy processes inside the crystal, the main mechanism being the polarization of the surrounding, usually highly polarizable, molecules. By realizing that organic molecules preserve their identity inside the crystal, we may estimate the screening energy, by describing the molecular interactions through classical electrostatics.

In BEDT-TTF and TTF-TCNQ families of organic conductors \cite{Ishiguro}, flat molecules crystallize forming low-dimensional anisotropic structures. Considering such a lattice of organic molecules, we ask what happens when a charge is added to one molecule. The electric field of that charge will induce a dipole moment in the surrounding highly polarizable molecules. Solving this electrostatic problem assuming fixed polarizabilities, we obtain the screening of the charged molecule in linear response. This gives the screening contribution of the lattice that is needed for renormalizing the bare Coulomb integral.

Modelling each molecule as a polarizable point with charge $q_i$, dipole moment ${\bf p}_i$ and polarizability $\alpha_i$ the energy of the lattice in the above situation is given by
\begin{eqnarray}
  \label{eq:Wq}
  W&=&\sum_{i} \left[ \frac{|{\bf {p}}_{i}|^2}{2{\alpha_i}}-{\bf {p}}_{i}\cdot{\bf {E}}_{i}^{ext}+
  \Phi_{i}q_{i}+T(q_{i}) \right]
  \nonumber \\
  && + \sum_{i,j>i} \left[ U_{1}({\bf p}_{i},{\bf p}_{j})+
  U_{2}({\bf p}_{i},q_j)+U_{3}(q_{i},q_{j}) \right],
\end{eqnarray}
where ${\bf E}_{i}^{ext}$ is the electric field and $\Phi({\bf r})$ the potential associated with the external
charge, while the first and fourth terms correspond to the resistance to polarization of each lattice point and to
the internal energy of the molecules as a function of charge state, respectively. The terms $U_{1}$, $U_{2}$ and
$U_{3}$ are the dipole-dipole, dipole-monopole and monopole-monopole interactions \cite{Jackson}. For molecules
without net charge (the electrons involved in charge transfer are explicitly included in the Hubbard model), we can
simplify Eq.~(\ref{eq:Wq}). Expressing it in the notation of Allen \cite{Allen}, defining the dipolar field by $|{\bf E}^{dip}\rangle=\Gamma |{\bf p}\rangle$, where $\Gamma$ is the dipole-dipole interaction matrix:
\begin{equation}
  \label{eq:Gammamatrix}
  \Gamma_{i\mu,j\nu} = \frac {3{r}_{ij\mu}{r}_{ij\nu}-\delta_{\mu \nu}{r}_{ij}^2}{r_{ij}^5}
  \quad,\quad i\neq j, \quad \mu,\nu \equiv x,y,z,
\end{equation}
the energy of the system is then given by
\begin{equation}
  \label{eq:WDirac}
  W=\frac {1}{2} \langle{{\bf p}| {\alpha}^{-1} - \Gamma |{\bf p}}\rangle-
  \langle{{\bf p}|{\bf E}^{ext}}\rangle.
\end{equation}
By applying the variational principle, $\delta W=0$, we obtain the expression for the set of dipoles which minimize the energy of the whole lattice,
\begin{equation}
  \label{eq:bracketp}
    |{\bf p}\rangle=  \left( {\alpha}^{-1} - \Gamma \right)^{-1} |{\bf E}^{ext}\rangle.
\end{equation}
Inserting Eq.~(\ref{eq:bracketp}) into Eq.~(\ref{eq:WDirac}) yields
\begin{equation}
  \label{eq:deltaW}
  \delta W = -\frac {1}{2} \langle{{\bf E}^{ext}| \left( {\alpha}^{-1} - \Gamma \right)^{-1}  |{\bf
   E}^{ext}}\rangle.
\end{equation}

Therefore, placing two charges at a lattice point or two different lattice points, we are able to calculate the correction to the total energy due to the polarization of the rest of the molecules in the crystal. Defining the external field as the composition of two fields, corresponding to charges placed at points $n$ and $m$ in the lattice, ${\bf E}^{ext}={\bf E}_n+{\bf E}_m$, the energy of the system is given by
\begin{equation}
  \delta W=-\langle{{\bf E}_m| \left( {\alpha}^{-1} - \Gamma
  \right)^{-1}|{\bf E}_n}\rangle- \langle{{\bf E}_n|  \left( {\alpha}^{-1} - 
  \Gamma \right)^{-1} |{\bf E}_m}\rangle.
\end{equation}
The equation above gives the on-site Coulomb screening, $\delta U$, when $n=m$, and the inter-site, $\delta V$, $\delta V'$, ..., for $n \neq m$. Thus, defining $|n-m|=l$, the general equation for the screening Coulomb parameters reads
\begin{equation}
  \label{eq:deltaVl}
  \delta V^l=-\langle{{\bf E}_0| \left( {\alpha}^{-1} - \Gamma \right)^{-1}|{\bf E}_l}\rangle\quad
  \Longrightarrow\quad  V^l=V_{0}^l+\delta V^l,
\end{equation}
where $\delta U$ corresponds to $l=0$, $\delta V$ to $l=1$, ... and $V_{0}^l$ are the Coulomb parameters for a single molecule/dimer, calculated with DFT \cite{TTF-TCNQ,Scriven}.

\begin{figure*}
\center
\vspace{3ex}
\includegraphics[scale=1.0]{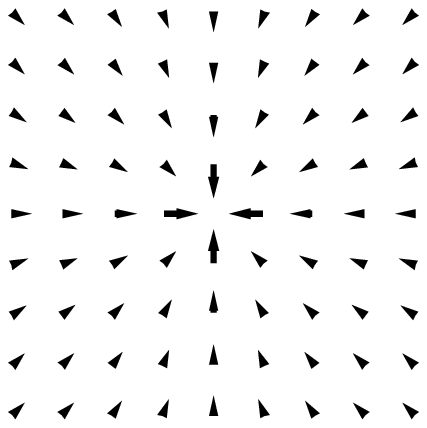}
\includegraphics[scale=1.0]{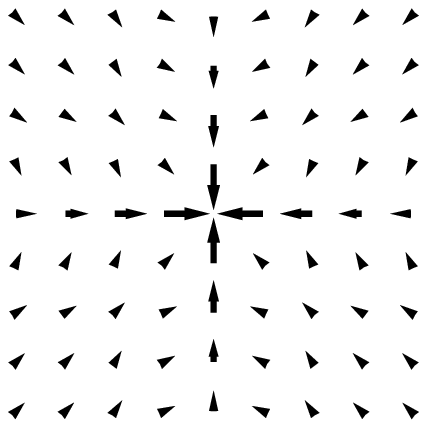}
\includegraphics[scale=1.0]{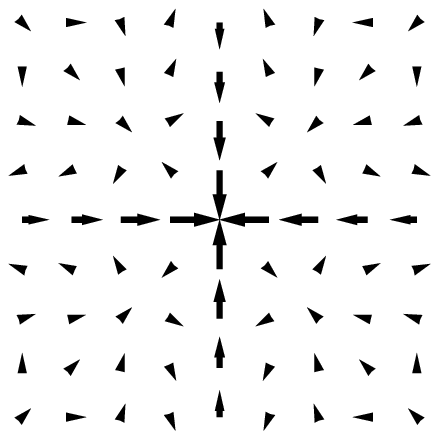}
\caption{Dipole arrangement in the $x$-$y$-plane of a simple-cubic lattice of point dipoles to a charge added at the center. As the polarizability increases from $\alpha=0.5 \alpha_{CM}$ (left), $0.75 \alpha_{CM}$ (middle) to $\alpha=0.8 \alpha_{CM}$ (right), the dipoles start to deviate strongly from pointing along the field-lines of the external field.}
\label{dipoleSC}
\end{figure*}

To illustrate the method we apply it to a cubic lattice of polarizable point dipoles, in
which an antiferroelectric instability occurs above a critical polarizability $\alpha_c$ 
\cite{Allen,Luttinger}, due to the anisotropy in the dipole interactions.
Such instability influences the screening of a charge in the lattice.
For small polarizabilities $\alpha$ the induced dipoles simply arrange along the field-lines of
the central point charge. Increasing $\alpha$ we, however, observe quite unconventional dipole arrangements as shown
in Fig.~\ref{dipoleSC}. This reflects the fact that for this lattice the matrix
$(\alpha^{-1}-\Gamma)$ becomes singular at a critical polarizability $\alpha_c$, with the eigenvector of the
vanishing eigenvalue having non-zero momentum (in this case ($\pi,\pi,0$)). Hence the ferroelectric instability happens before the critical value $\alpha_{CM}$ obtained from the Clausius-Mossotti relation \cite{Allen}.

\section{Submolecular approach}
\label{sec:submolecular}
In a real molecular crystal the molecules typically form stacks, i.e., they are quite closely packed. Consequently,
approximating the polarizability of a molecule by just one point-polarizability is not a good approximation. It only
works outside a sphere containing the charge density of the molecule, which for typical $\pi$-bonded molecules will
intersect the neighboring molecules, but it becomes accurate at large distances. To obtain a better description of
the molecular response, we model it as point-polarizabilities distributed over the non-hydrogen atoms of the
molecule \cite{Mazur}. The accuracy of this submolecular approach is tested for TTF-TCNQ salts.

We consider a small system of only three or five molecules in a TTF chain. For such small systems we can perform constrained density-functional calculations \cite{TTF-TCNQ,Behler}, constraining the added charge to the central molecule. This gives us the screening due to the molecules in this small assembly. We compare to the results of the screening calculated using the distributed point-polarizability approach. Since this approach works least well in the near-field, this is a critical test of the method. Nevertheless, both approaches agree within $5\%$. If we included more distant molecules, for which the computational cost of constrained density-functional calculations would be prohibitive, the agreement would improve further, as the multipole approximation becomes exact in the far-field. Thus, using polarizability tensors obtained with DFT \cite{TTF-TCNQ} in our submolecular approach essentially reproduces the results of full quantum mechanical treatment of the screening.

\section{Screening parameters in TTF-TCNQ}
\label{sec:screening}
\begin{figure}
\center
\includegraphics[scale=0.35,clip]{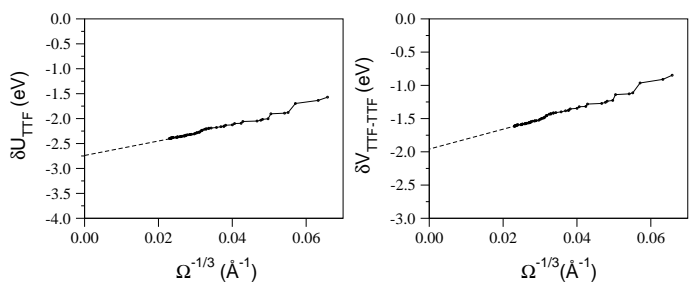}
\caption{Screening of on-site Coulomb interaction, $\delta U$, in a TTF molecule (left panel), and screening of
inter-site Coulomb interaction, $\delta V$, for nearest-neighbor TTF molecules (right panel), versus
$\Omega^{-1/3}$, for clusters of increasing radius. Extrapolation to the thermodynamic limit is shown by the dotted
line.}
\label{UTTF}
\end{figure}

To obtain the screened parameters for a TTF-TCNQ crystal, we perform calculations for crystal fragments (clusters)
of increasing sizes to extrapolate to the infinite-size limit. To obtain the parameters for the extended Hubbard
model (\ref{eq:HubbardModel}) we consider the Coulomb screening between two charges at the same or at neighbor
molecules placed in the center of the cluster. For simplicity, the clusters of $N$ molecules are constructed as
spheres of radius $R$, where $R$ is the distance between the doped molecule (or center of the pair of molecules) and the farthest
neighbor in the cluster, hence $\Omega=\frac {4\pi}{3}R^3$ being the volume of the sphere \cite{Pederson}. For large
enough $R$ we can linearly extrapolate the screening energy $\delta U$ or $\delta V$, respectively, versus $1/R$ to
the thermodynamic limit ($R \to \infty$). This is shown in Fig.~\ref{UTTF}, where clusters of up to 400 molecules
are studied. Calculations for TTF-TCNQ salts up to the third nearest-neighbors are performed and results are given
in Table \ref{screening}. The ratio of the radius of the sphere, $R$, to the distance between
third nearest-neighbors, $d$, where we still can make the extrapolation to the thermodynamic limit is 
$d/R = 0.43$. However for longer range interactions, the cluster size should be increased 
towards this convergence ratio, reaching a linear dependence with the inverse of the volume, which
allows the extrapolation to the bulk value.

\begin{table}
\center
\begin{tabular}{l@{\hspace{3ex}}r@{\hspace{3ex}}r@{\hspace{3ex}}r@{\hspace{3ex}}r}
      & $\delta U$ & $\delta V$ & $\delta V'$ & $\delta V''$\\ \hline
 TTF  & 2.7           & 1.9           & 1.25     &0.8\\
 TCNQ & 2.6           & 1.9           & 1.3      &0.9\\
\end{tabular}
\caption{\label{screening}
Screening Coulomb energies for TTF-TCNQ.
$\delta U$ is the on-site screening and $\delta V$, $\delta V'$ and $\delta V''$
correspond to first, second and third nearest-neighbor screening energies between molecules
in the same TTF or TCNQ stack. All energies are in eV. For comparison, the
band-width is about 0.7~eV.}
\end{table}

\section{Conclusions}
\label{sec:conclusions}
We have presented a general method for obtaining accurate values of Coulomb parameters in organic molecular
compounds, where molecules interact weakly, and shown how it works for TTF-TCNQ.
We are currently extending this method to other crystals like
quasi-bidimensional salts based on the BEDT-TTF molecule.

\section*{Acknowledgements} 

JM and LC acknowledge financial support from MICINN under contract
CTQ2008-06720-C02-02.



\end{document}